%% file: SPAWC2017_v3.tex
\documentclass[conference, twocolumn]{IEEEtran-v17}

\usepackage{xspace}
\usepackage{url}
\usepackage[cmex10]{amsmath}
\usepackage{bbm}
\usepackage{graphicx}
\usepackage{epstopdf}
\usepackage{paralist}
\usepackage[normalem]{ulem}
\usepackage{fancyref} 
\usepackage{amsmath}
\usepackage{amssymb}

\usepackage{dsfont}
\usepackage{footnote}
\usepackage{color}
\usepackage{balance}
\usepackage{float}
\usepackage{cite}
\usepackage{wasysym}     
\usepackage{amssymb}

\usepackage{vmr-symbols-vecbold}
\usepackage{standard-macros}
\usepackage{mathbbol}

\usepackage{mdframed}
\newmdtheoremenv{theo}{Theorem}
\newmdtheoremenv{coro}{Corollary}
\newmdtheoremenv{lemo}{Lemma}

\input{./symbols}

\usepackage{pgfplots}
\usepgfplotslibrary{external}
\tikzset{external/system call={latex \tikzexternalcheckshellescape -halt-on-error
-interaction=batchmode -jobname "\image" "\texsource" &&
dvips -o "\image".ps "\image".dvi}}
\pgfkeys{/pgf/images/include external/.code=\includegraphics{#1}}

\begin{document}

\IEEEoverridecommandlockouts
\title{Finite-Blocklength Bounds on the Maximum Coding Rate of Rician Fading Channels with Applications to Pilot-Assisted Transmission}

 \author{\IEEEauthorblockN{Johan \"Ostman, Giuseppe Durisi, and Erik G. Str\"om\\
Dept. of Signals and Systems, Chalmers University of Technology,  Gothenburg, Sweden}
 \thanks{To appear in Proceedings of the IEEE International Workshop on Signal Processing Advances in Wireless Communications (SPAWC 2017).}}
\maketitle

\begin{abstract}
  We present nonasymptotic bounds on the maximum coding rate achievable over a Rician block-fading channel for a fixed packet size and a fixed packet error probability.
  Our bounds, which apply to the scenario where no \emph{a priori} channel state information is available at the receiver, allow one to quantify the tradeoff between the rate gains resulting from the exploitation of time-frequency diversity and the rate loss resulting from fast channel variations and pilot-symbol overhead. 
\end{abstract}

\section{Introduction} 
\label{sec:introduction}

To enable future autonomous systems such as connected vehicles, automated factories, and smart grids, next-generation wireless communication systems must be able to support the sporadic transmission of short data packets within stringent latency and reliability constraints~\cite{metis-project-deliverable-d1.113-04a, durisi16-09a}.
Classic information-theoretic performance metrics such as \emph{ergodic} and  \emph{outage capacity}, provide inaccurate benchmarks to the performance of short-packet coding schemes, because of their asymptotic nature~\cite{durisi16-09a,durisi16-02a}. 
In particular, these performance metrics are unable to capture the tension between the throughput gains in the transmission over wireless fading channels  attainable by exploiting channel diversity and the throughput losses caused by the insertion of pilot symbols, which are needed to estimate the wireless fading channel (\emph{pilot overhead}).

In this paper, we provide a characterization of the tradeoff between latency, reliability, and throughput in the transmission of short packets over point-to-point \emph{Rician} block-fading channels. 
Our analysis explicitly accounts for the pilot overhead.

%
\paragraph*{Relevant prior art} 
\label{par:relevant_prior_art}
The fundamental quantity of interest in short-packet communications is the \emph{maximum coding rate} $\Rmax(n,\epsilon)$, which is  the largest rate achievable by any channel code having  blocklength $n$ and packet error probability no larger than $\epsilon$. 
Note that the classic Shannon capacity can be obtained from $\Rmax(n,\epsilon)$ by taking the limit $n\to\infty$ and $\epsilon\to 0$.

No closed-form expressions for~$\Rmax(n,\epsilon)$ are available for the channel models of interest in wireless communication systems.
However, tight numerically computable bounds on~$\Rmax(n,\epsilon)$ have been recently  obtained for a variety of channels; such bounds rely on the nonasymptotic tools recently developed by Polyanskiy, Poor, and Verd\'u~\cite{polyanskiy10-05a}.
We next summarize the available results, starting with the nonfading \emph{complex} AWGN channel. 
For this channel,  tight upper (converse) and lower (achievability) bounds on $\Rmax(n,\epsilon)$ based on cone packing were obtained by Shannon~\cite{shannon59-a}.
Polyanskiy, Poor, and Verd\'u~\cite{polyanskiy10-05a}  showed recently that Shannon's converse bound is a special case of the so-called \emph{min-max converse}~\cite[Th.~27]{polyanskiy10-05a}, \cite{polyanskiy13-07b}, a general converse bound that involves a binary hypothesis test between the channel law and a suitably chosen auxiliary distribution. 
Furthermore, they obtained an alternative achievability bound---the $\kappa\beta$-bound~\cite[Th.~25]{polyanskiy10-11a}---also based on binary hypothesis testing.
This bound, although less tight than Shannon's achievability bound, is easier to evaluate numerically and to analyze asymptotically. 
Indeed, Shannon's achievability bound relies on the transmission of codewords that are uniformly distributed on the surface of an $n-1$-dimensional complex hypersphere in $\complexset^n$ (a.k.a., \emph{spherical} or \emph{shell codes}), which makes the induced output distribution unwieldy. 
Min-max and $\kappa\beta$ bounds solve this problem by replacing this output distribution by a product Gaussian distribution.

Analyzing the min-max converse and the $\kappa\beta$ bound in the asymptotic regime of large blocklength $n$, Polyanskiy, Poor, and Verd\'u established the following asymptotic expansion for $\Rmax(n,\epsilon)$ (see~\cite{polyanskiy10-05a} and also the refinement in~\cite{tan15-05a}):
\IEEEpubidadjcol
\begin{IEEEeqnarray}{rCL}\label{eq:normal_approximation}
  \Rmax(n,\epsilon)= C- \sqrt{n^{-1}V}Q^{-1}(\epsilon) +\landauO\lefto({n^{-1}\log n}\right).
\end{IEEEeqnarray}
Here, $C=\log(1+\rho)$, where $\rho$ denotes the SNR, is the channel capacity, $V=\rho(2+\rho)/(1+\rho)^2$ is the so-called channel \emph{dispersion}, $Q(\cdot)$ is the Gaussian $Q$ function, and $\landauO(n^{-1}\log n)$ comprises reminder terms of order $n^{-1}\log n$.

We next move to the fading case and focus  on the setup where no \emph{a priori} channel-state information (CSI) about the fading channel is available at transmitter and receiver. 
The assumption of no \emph{a priori} CSI at the receiver is of particular relevance for short-packet transmission because information-theoretic analyses conducted under this assumption account automatically for the ``cost'' of acquiring CSI~\cite{lapidoth05-07a,yang13-02a,devassy15-07a}.
Bounds on $\Rmax$ for generic \emph{quasi-static multiple-antenna fading channels} were reported in~\cite{yang14-07c}.
Using these bounds, the authors showed that under mild condition on the fading distribution, the channel dispersion (i.e., the term $V$ in~\eqref{eq:normal_approximation}) is zero. 
This means that the asymptotic limit (in this case the outage capacity) is approached much faster in $n$ than in the AWGN case.
This is because the main source of error in quasi-static fading channels is the occurrence of ``deep fades''; channel codes cannot mitigate them.
The achievability bound in~\cite{yang14-07c} relies on a modified version of the $\kappa\beta$ bound, where the decoder computes the angle between the received signal and each one of the codewords.
The converse bound relies on the min-max converse.
The analysis in~\cite{yang14-07c} was later partly generalized in~\cite{durisi16-02a} to fading channels offering time-frequency diversity.
Specifically, the authors of~\cite{durisi16-02a}  focused on a \emph{multi-antenna Rayleigh block-fading model} where coding is performed across a fixed number of coherence blocks. 
Their converse bound relies again on the min-max converse, whereas the achievability bound relies on the so-called dependency-testing (DT) bound~\cite[Th.~17]{polyanskiy10-05a}. 
The input distribution used in the DT bound is the one induced by \emph{unitary space-time modulation} (USTM), where the  matrices describing the signal transmitted within each coherence block  are (after a normalization) uniformly distributed on the set of unitary matrices. 
This distribution, which achieves capacity at high SNR~\cite{zheng02-02a,yang13-02a}, coincides with the one induced by shell codes in the single-input single-output (SISO) case.
The auxiliary distribution used in the min-max converse is the one induced by USTM.
Unfortunately, this distribution is unwieldy.
As a consequence, no
asymptotic expansions for $\Rmax$ similar to~\eqref{eq:normal_approximation} are available.

\paragraph*{Contributions} 
\label{par:contributions}
We provide upper and lower bounds on $\Rmax$ 
for \emph{SISO Rician block-fading channels} under the assumption of no \emph{a priori} CSI. 
Similar to~\cite{durisi16-02a}, our bounds rely on the min-max converse and the DT bound, and on the transmission of shell codes.
The bounds recover the ones obtained in~\cite{durisi16-02a} for the Rayleigh fading when the Rician factor  $\kappa$ is set to $0$, and agree with the normal approximation~\eqref{eq:normal_approximation} when $\kappa\to \infty$.

We also provide an extension of our achievability bound to the case when pilot symbols are used  to estimate the channel at the decoder.
Our analysis provides a nonasymptotic perspective on the pilot-assisted transmission problem, which has been addressed so far only in the asymptotic regime of large packet size (see, e.g.,~\cite{hassibi03-04a,tong04-11a,jindal09-06a}).

\paragraph*{Notation}
Uppercase letters such as $X$ and $\randvecx$ are used to denote scalar random variables and vectors, respectively, and the realizations are written in lowercase, e.g., $x$ and $\vecx$.
The identity matrix of size $a\times a$ is written as $\matI_{a}$.
The distribution of a circularly-symmetric complex Gaussian random variable with variance $\sigma^2$ is denoted by $\cgauss{0}{\sigma^2}$. 
The superscript~$\tp{}$ denotes transposition, $\herm{}$ Hermitian transposition, and $\odot$ the Schur product. 
Furthermore, $\veczero_{n}$ and $\vecone_{n}$ stand for the all-zero and all-one vectors of size $n$, respectively.
Finally, $\log\lro{\cdot}$ indicates the natural logarithm, $\lrh{a}^+$ stands for $\max\lrbo{0, a}$, $\Gamma\lro{\cdot}$ denotes the Gamma function, $I_{\nu}\lro{z}$ the modified Bessel function of the first kind, $\vecnorm{\cdot}$ the $l^2$-norm, and $\Ex{}{\cdot}$  the expectation operator.

\section{System Model} 
\label{sec:system_model}
We consider a single-input single-output Rician block-fading channel.
Specifically, the random non-line-of-sight (NLOS) component is assumed to stay constant for $\nc$ successive channel uses (which form one coherence block) and  to change independently across coherence blocks.
Coding is performed across $\ell$ such blocks; we shall refer to $\ell$ as the number of \emph{time-frequency diversity branches}.
The duration of each codeword (packet size) is, hence, $n=\nc\ell$.
The LOS component, which is assumed to be known at the receiver, stays constant over the duration of the entire packet (codeword).
No \emph{a priori} knowledge of the NLOS component is available at the receiver, in accordance to the no-CSI assumption.
Mathematically, the channel input-output relation can be expressed as
\begin{IEEEeqnarray}{rCl}
\label{eq:sys_mod}
\randvecy_k =  H_k\vecx_k + \randvecw_k, \quad k = 1,\dots, \ell.
\end{IEEEeqnarray}
Here, $\vecx_k \in \complexset^{\nc}$, $\randvecy_k \in \complexset^{\nc}$ are vectors containing the transmitted and received symbols within block $k$, respectively, and $H_k \distas \cgauss{\mu\sub{H}}{\sigma^2\sub{H}}$ is the Rician-fading coefficient. 
Here, $\mu\sub{H}=\sqrt{\kappa/(1+\kappa)}$ and  $\sigma^2\sub{H} = (1+\kappa)^{-1}$ where $\kappa$ is the Rician factor. 
Finally, the vector $\randvecw_k\distas \cgauss{\bm{0}}{\matI_{\nc}}$ models the AWGN process. 
The random variables $\lrb{H_k}$ and $\lrb{\randvecw_k}$, which are mutually independent, are also independent over $k$. 

We next define a channel code.
\begin{dfn}
An $\lro{\ell, \nc, M, \epsilon, \rho}$-code for the channel (\ref{eq:sys_mod}) consists of
\begin{itemize}
\item 
An encoder $f:\lrbo{1, \dots, M} \rightarrow  \complexset^{\nc\ell}$ that maps the message $J\in \lrbo{1,\dots, M}$ to a codeword in the set $\lrb{\vecc_1,\dots, \vecc_M}$.
Since each codeword $\vecc_m$, $m=1\dots, M$, spans $\ell$ blocks, it is convenient to express it as a concatenation of $\ell$ subcodewords of dimension $\nc$
\begin{IEEEeqnarray}{rCl}
	\vecc_m = \lrho{\vecc_{m,1}, \dots, \vecc_{m,\ell}}.
\end{IEEEeqnarray}
We require that each subcodeword satisfies the average-power constraint
\begin{IEEEeqnarray}{rCl}\label{eq:power_constraint}
\vecnorm{\vecc_{m,k}}^2 =  \nc\rho, \quad k=1,\dots, \ell.
\end{IEEEeqnarray}
Since the noise has unit variance, we can think of $\rho$ as the SNR.

\item A decoder $g: \complexset^{\nc\ell} \rightarrow \lrbo{1,\dots, M}$ satisfying an average error probability constraint
\begin{IEEEeqnarray}{rCl}\label{eq:avPrC}
\frac{1}{M} \sum_{j=1}^M \Pr\lrbo{g\lro{\randvecy^\ell} \neq J \given J=j}\leq \epsilon
\end{IEEEeqnarray}
where $\randvecy^\ell = \lrho{\randvecy_1, \dots, \randvecy_\ell}$ is the channel output induced by the codeword $\vecx^\ell = \lrho{\vecx_1, \dots, \vecx_\ell} = f(j)$.
\end{itemize}
\end{dfn}
For given $\ell$ and $n_c$, $\epsilon$, and $\snr$, the maximum coding rate \Rmax, measured in information bits per channel use, is defined as follows:
\begin{IEEEeqnarray}{rCl}
\label{eq:maximal_rate}
R^* \triangleq \sup\lrbo{\frac{\log_2\lro{M}}{\ell\nc} : \exists \lro{\ell, \nc, M, \epsilon, \rho}\text{--code}}.
\end{IEEEeqnarray}
\section{Finite-blocklength bounds on \Rmax} 
\subsection{An Auxiliary Lemma} 
\label{sec:an_auxiliary_lemma}

We next present our achievability and converse bounds on $\Rmax$ in~\eqref{eq:maximal_rate}. 
The achievability bound relies on the DT bound~\cite[Th.~17]{polyanskiy10-05a} and on the transmission of independent shell codes over each coherence block. 
This achievability bound does not require the explicit estimation of the fading coefficients; rather, it relies on a \emph{noncoherent} transmission technique in which the message is encoded in the direction of each vector $\vecx_k$ in~\eqref{eq:sys_mod}--a quantity that is not affected by the fading process. 
The case of explicit channel estimation through pilot-assisted transmission will be treated in Section~\ref{sec:pilots}.

Our converse bound relies on the min-max converse~\cite[Th.~27]{polyanskiy10-05a}, with auxiliary distribution chosen as the one induced on $\lrbo{\randvecy_k}$ by the transmission of independent shell codes over each coherence block.
We start by providing in the next lemma the output distribution induced by a shell code of length $n_c$.
Its proof is omitted for space constraints.
\begin{lem}
\label{lem:ustm_induced_pdf}
Let $\randvecx \in \complexset^{\nc}$ be uniformly distributed on the $(\nc-1)$--dimensional complex hypersphere of radius $\sqrt{\rho \nc}$ and let $H\sim \cgauss{\mu}{\sigma^2}$. Furthermore, let $\randvecy = H\randvecx  + \randvecw$ where $\randvecw$ is defined as in (\ref{eq:sys_mod}). The probability density function (pdf) of $\randvecy$ is given by
\begin{IEEEeqnarray}{rCl}\label{eq:shell_induced_out_pdf}
  f_{\randvecy}\lro{\vecy} &&= \frac{\Gamma\lro{\nc}}{\sigma^2 \pi^{\nc}} e^{-\vecnorm{\vecy}^2} e^{-\frac{\abs{\mu }^2}{\sigma^2}} 
 \int_{\positivereals} \frac{e^{-\lro{\rho\nc + \sigma^{-2}} z }}{ \lro{\sqrt{\vecnorm{\vecy}^2 \rho \nc z}}^{\nc-1} } \nonumber\\
&&\times I_{0}\lro{2\sqrt{{z\abs{\mu}^2 }/{\sigma^4}}}  I_{\nc-1}\lro{2\sqrt{\vecnorm{\vecy}^2 \rho \nc z}} \mathrm{d}z.
\end{IEEEeqnarray}
\end{lem}
%

\subsection{A Noncoherent Lower Bound on \Rmax} 
\label{sec:a_noncoherent_lower_bound_on_rmax}

We are now ready to state our lower bound on \Rmax.
\begin{thm}[DT lower bound]
\label{thm:DT}
\Rmax in~\eqref{eq:maximal_rate} is lower-bounded as
\begin{IEEEeqnarray}{rCl}
	R^* \geq \max\lrbo{\frac{\log_2\lro{M}}{\nc\ell} : \epsilon\sub{ub}\lro{M} \leq \epsilon}
\end{IEEEeqnarray}
where 
\begin{equation}
	 \epsilon\sub{ub}\lro{M} = \Ex{}{\exp\lefto\{-\lrho{ \sum_{k=1}^\ell S_k - \log\lro{\frac{M-1}{2}} }^+\right\}}
\end{equation}
with 
\begin{IEEEeqnarray}{rCl}
\label{eq:Sk}
	 S_k &=& \frac{\abs{\mu\sub{H}}^2}{\sigma\sub{H}^2}  -  \vecnorm{\randvecw_k}^2 - \log\lro{\sigma\sub{H}^2\lro{\sigma\sub{H}^2\nc\rho + 1}} \nonumber \\
	  &-&\:\log \int_{\positivereals} \frac{e^{-\lro{\rho\nc + \sigma\sub{H}^{-2}} z }}{ \lro{\sqrt{\vecnorm{\widetilde{\randvecw}_k}^2 \rho \nc z}}^{\nc-1} } \nonumber\\
	& &  \hspace*{-1cm} \times \: I_{0}\lro{2\sqrt{{z\abs{\mu\sub{H}}^2}/{\sigma\sub{H}^4} }} I_{\nc-1}\lro{2\sqrt{\vecnorm{\widetilde{\randvecw}_k}^2 \rho \nc z}} \mathrm{d}z.
\end{IEEEeqnarray}
Here, $\randvecw_k$ is defined as in \eqref{eq:sys_mod} and
\begin{IEEEeqnarray}{rCl}
\widetilde{\randvecw}_k =\begin{bmatrix} \mu\sub{H} \sqrt{\nc\rho} \\ \veczero_{\nc-1} \end{bmatrix} +\begin{bmatrix} \sqrt{\sigma\sub{H}^2 \nc \rho + 1} \\ \vecone_{\nc-1} \end{bmatrix} \odot \randvecw_k.
\end{IEEEeqnarray}
\end{thm}

\begin{IEEEproof}
The proof follow steps similar to the ones reported in \cite[App. A]{durisi16-02a}.
Specifically, we let $\randvecx_k = \sqrt{\nc \rho} \randvecu_k$ where $\lrb{\randvecu_k}_{k=1}^\ell$ are independent and isotropically distributed unitary vectors.
It follows from Lemma~\ref{lem:ustm_induced_pdf} that the vectors $\randvecy_k = \sqrt{\nc \rho} \randvecu_k H_k + \randvecw_k$,  $k\in \lrb{1,\dots, \ell}$, are independent and $f_{\randvecy}$-distributed. 

The block-memoryless assumption implies that the information density \cite[Eq.~(4)]{polyanskiy10-05a} can be decomposed as
\begin{equation}
\label{eq:info_dens}
i\lro{\vecu^\ell; \vecy^\ell} = \sum_{k=1}^\ell i\lro{\vecu_k; \vecy_k} = \sum_{k=1}^\ell \log\frac{f_{\randvecy \given \randvecu}\lro{\vecy_k \given \vecu_k}}{ f_{\randvecy}\lro{\vecy_k}}
\end{equation}
where 
\begin{equation}\label{eq:cond_distr_unit}
  f_{\randvecy\given \randvecu=\vecu_k}=\cgauss{ \mu\sub{H} \sqrt{\nc\rho} \vecu_k}{ \sigma\sub{H}^2\nc\rho \vecu_k \herm{\vecu}_k + \matI_{\nc} }
\end{equation}
and $f_{\randvecy}$ is given in~\eqref{eq:shell_induced_out_pdf}.
One can also verify that for every $\nc\times\nc$ unitary matrix $\matV$,
\begin{IEEEeqnarray}{rCl}
\label{eq:cond_indep_x}
f_{\randvecy \given \randvecu}\lro{\vecy_k \given \herm{\matV}\vecu_k} = f_{\randvecy \given \randvecu}\lro{\matV \vecy_k \given \vecu_k}
\end{IEEEeqnarray}
and
\begin{IEEEeqnarray}{rCl}
\label{eq:induced_indep_x}
 f_{\randvecy}\lro{\matV \vecy_k } = f_{\randvecy}\lro{\vecy_k }.
\end{IEEEeqnarray}
This implies that $i\lro{\vecu_k;\randvecy_k}$ does not depend on $\vecu_k$ when $\randvecy_k \sim f_{\randvecy}$. 
Hence, we can set without loss of generality $\vecu_k = \tp{\lrho{1,0,\dots,0}}$, $k = 1,\dots, \ell$.
One can finally show that, when $\randvecy_k \sim f_{\randvecy\given \randvecu=\vecu_k}$, the information density $i\lro{\vecu_k; \randvecy_k}$ has the same distribution as the random variable $S_k$ in \eqref{eq:Sk}.
The proof is concluded by invoking the DT bound~\cite[Th. 17]{polyanskiy10-05a}.
\end{IEEEproof}

\subsection{An Upper Bound on \Rmax} 
\label{sec:a_general_upper_bound_on_rmax}


We next state our converse bound.
\begin{thm}[Min-max converse bound]
\label{thm:MC}
\Rmax in~\eqref{eq:maximal_rate} is upper-bounded as
\begin{equation}
\label{eq:MC}
 R^*\leq \inf_{\lambda \geq 0} \frac{1}{\ell\nc} \lro{\lambda - \log\lrho{\Pr\lrbo{\sum_{k=1}^\ell S_k \leq \lambda} - \epsilon}^+}
\end{equation}
where the $\{S_k\}$ are defined in~\eqref{eq:Sk}.
\end{thm}
\begin{IEEEproof}
 We use as auxiliary channel in the min-max converse~\cite[Th.~27]{polyanskiy10-05a}, the one for which $\vecy^\ell$ has pdf
  \begin{IEEEeqnarray}{rCl}
  	q_{\randvecy^\ell}\lro{\vecy^\ell} = \prod_{k=1}^{\ell} f_{\randvecy}\lro{\vecy_k}
  \end{IEEEeqnarray}
where $f_{\randvecy}$ is given in~\eqref{eq:shell_induced_out_pdf}.
For this choice, it follows from (\ref{eq:Sk}), (\ref{eq:cond_indep_x}), and (\ref{eq:induced_indep_x}) that the Neyman-Pearson function $\beta\lro{\vecx^\ell, q_{\randvecy^\ell}}$ defined in \cite[Eq. (105)]{polyanskiy10-05a} is independent of $\vecx^\ell$.
Hence, we can use~\cite[Th. 28]{polyanskiy10-05a} to conclude that 
\Rmax is upper-bounded as
\begin{IEEEeqnarray}{rCl}
\label{eq:MC2}
	\Rmax \leq \frac{1}{\nc\ell} \log\frac{1}{\beta_{1-\epsilon} \lro{\vecx^\ell, q_{\randvecy^\ell}}}.
\end{IEEEeqnarray}
Without loss of generality, we shall set $\vecx_k=[\sqrt{n_c\rho},0\dots,0]$, $k=1,\dots,\ell$.
It follows by the Neyman-Pearson lemma~\cite{neyman33-01a} that 
\begin{equation}
	\beta_{1-\epsilon} \lro{\vecx^\ell, q_{\randvecy^\ell}} = \Pr\lrbo{r\lro{\vecx^\ell; \randvecy^\ell} \geq \gamma}, \quad \randvecy^\ell \sim q_{\randvecy^\ell} 
\end{equation}
where $\gamma$ is the solution to
\begin{equation}
	\Pr\lrbo{r\lro{\vecx^\ell; \randvecy^\ell} \leq \gamma} = \epsilon, \quad \randvecy^\ell \sim f_{\randvecy^\ell \given \randvecx^\ell}
\end{equation}
and
\begin{equation}
\label{eq:info_dens_miss}
r\lro{\vecx^\ell; \vecy^\ell} =\sum_{k=1}^\ell r\lro{\vecx_k; \vecy_k}=\sum_{k=1}^\ell \log\frac{f_{\randvecy \given \randvecx}\lro{\vecy_k \given \vecx_k}}{ f_{\randvecy}\lro{\vecy_k}}.
\end{equation}
Finally, we obtain~\eqref{eq:MC} by relaxing~\eqref{eq:MC2} using~\cite[Eq. (106)]{polyanskiy10-05a} (which yields a generalized Verd\'u-Han converse bound, cf.~\cite{verdu94-07a}) and by exploiting that when $\randvecy_k \distas f_{\randvecy \given \randvecx=\vecx_k}$ the random variable $r\lro{\vecx_k; \randvecy_k}$ is distributed as $S_k$ in~\eqref{eq:Sk}.
\end{IEEEproof}

\paragraph*{Remark} 
The achievability and converse bounds reported in Theorem~\ref{thm:DT} and~\ref{thm:MC} coincide with the bounds obtained in~\cite{durisi16-02a} for the Rayleigh-fading case if one sets $\kappa=0$ and replaces the maximum probability of error constraint used in~\cite{durisi16-02a} with the average probability of error constraint~\eqref{eq:avPrC}.
\subsection{A Pilot-Assisted Lower Bound on \Rmax} \label{sec:pilots}
We next present a lower bound on $\Rmax$ for the case in which pilot symbols are transmitted to enable the decoder to perform channel estimation.
Specifically, we  assume that within each coherence block, $\np$ out of the available $\nc$ channel uses are reserved for pilot symbols.
The remaining $\nd=\nc-\np$ channel uses are left for data symbols.
We  further assume that all pilot symbols are transmitted at power $\rho$, and that each data symbol vector $\vecx_k^{(\text{d})}\in \complexset^{\nd}$, $k=1,\dots,\ell$  satisfies the power constraint $\vecnorm{\vecx_k^{(\text{d})}}^2=\nd \rho$ so that \eqref{eq:power_constraint} holds.

The receiver uses the $\np$ pilot symbols per coherence block to perform a maximum likelihood estimate of the fading coefficient within the coherence block. 
Specifically, given $H_k=h_k$, the receiver obtains the estimate $\widehat{H_k} \sim \cgauss{h_k}{\sigma\sub{e}^2}$ where $\sigma\sub{e}^2 = (\np \rho)^{-1}$.
This implies that, given the channel estimates $\{\widehat{H}_k=\widehat{h}_k\}$, $k=1,\dots,\ell$ (which are available at the receiver), we can  express the input-output relation for the data symbols in the following equivalent form:
\begin{equation}\label{eq:io_relation_pilots}
  	\randvecy_k = Z_k\vecx_k  + \randvecw_k, \quad k=1,\dots,\ell.
\end{equation}
Here, all vectors belong now to $\complexset^{\nd}$ and the random variable $Z_k$ is
 $\cgauss{\mu\sub{p}(\widehat{h}_k)}{\sigma\sub{p}^2}$-distributed with
\begin{equation}
  \mu\sub{p}(\widehat{h}_k) = \frac{\sigma\sub{H}^2 \widehat{h}_k + \sigma\sub{e}^2 \mu\sub{H}}{\sigma\sub{H}^2+\sigma\sub{e}^2}, \quad \sigma\sub{p}^2 = \frac{\sigma\sub{H}^2\sigma\sub{e}^2}{\sigma\sub{H}^2+\sigma\sub{e}^2}.
\end{equation}
We see from~\eqref{eq:io_relation_pilots} that we can account for the availability of the noisy CSI $\{\widehat{H}_k=\widehat{h}_k\}$ simply by transforming the Rician fading channel~\eqref{eq:sys_mod} into the equivalent Rician fading channel~\eqref{eq:io_relation_pilots}, whose LOS component is a random variable that depends on the channel estimates $\{\widehat{H}_k\}$.
A lower bound on $\Rmax$ for this setup can be readily obtained by assuming that each $\nd$ dimensional data vector is generated independently from a shell code, by applying Theorem~\ref{thm:DT} to each realization of $\{\widehat{H}_k\}$, and then by averaging over $\{\widehat{H}_k\}$. 
The resulting bound is given in Theorem~\ref{thm:pilot} below.
\begin{thm}[Pilot-assisted DT lower bound]\label{thm:pilot}
Assume that $\np$ pilots per coherence interval are used to estimate the fading coefficients.
Then \Rmax in~\eqref{eq:maximal_rate} is lower-bounded as 
\begin{IEEEeqnarray}{rCl}
	R^* \geq \max\lrbo{\frac{\log_2\lro{M}}{\nc\ell} : \epsilon^{(\np)}\sub{ub}\lro{M} \leq \epsilon}
\end{IEEEeqnarray}
where 
\begin{IEEEeqnarray}{rCl}\label{eq:error_pilot}
	&&   \hspace*{-0.5cm}\epsilon^{(\np)}\sub{ub}\lro{M}= \nonumber \\
	 && \Ex{}{\exp\lefto\{-\lrho{ \sum_{k=1}^\ell \bar{S}_k(\widehat{H}_k) - \log\lro{\frac{M-1}{2}} }^+\right\}}.
\end{IEEEeqnarray}
Note that the expectation in~\eqref{eq:error_pilot} is computed also with respect to the channel estimates $\{\widehat{H}_k\}$; the random variables $\{\bar{S}_k(\widehat{H}_K)\}$ are defined similarly as in~\eqref{eq:Sk} with the difference that $\nc$, $\mu\sub{H}$ and $\sigma\sub{H}^2$ in~\eqref{eq:Sk} are replaced by $\nd$, $\mu\sub{p}(\widehat{H_k})$ and $\sigma\sub{p}^2$, respectively.
\end{thm}
\paragraph*{Remark} 
For the case $\np=0$, the pilot-based achievability bound in Theorem~\ref{thm:pilot} coincides with the noncoherent bound given in Theorem~\ref{thm:DT}.

\section{Numerical Results} 
\label{sec:numerical_results}

\subsection{Dependency of \Rmax on the Rician Factor $\kappa$ } 
\label{sec:numerical_results_a}

\begin{figure}[h]
 \centering
\begin{tikzpicture}
\begin{axis}
    [
    	axis x line*=bottom,
    	xmode = log,        
         xlabel={Number of time-frequency diversity branches $\ell$ (log scale)},
        ylabel = Bit/channel use,
        xmin = 2,
        xmax = 84,
        ymin = 0,
        xticklabels={{$2$}, {$4$}, {$7$}, {$14$}, {$21$},{$28$}, {$42$},{$84$}},
       xtick={2, 4 , 7,14,21,28,42, 84},   
		legend style={at={(axis cs:5,0)},anchor=south west},
		 xlabel near ticks,
		 grid=both
    ]
		 \def\ptsize{1.5}
      
        \addplot[color = black, ,mark size=\ptsize,line width=0.5mm] table [y index={1}, col sep=comma] {./Data/Gaussian_ref.csv}coordinate[pos=0.28](p1);      
        \coordinate (p2) at ($(p1)+ (-3pt,7pt)$); \draw[<-] (p1)--(p2) node at ($(p2) + (-10pt,3pt)$){Normal Approximation};
        		
		 \addplot[name path = p1,color = blue, solid,mark=*,mark size=\ptsize,forget plot] table [y index={1}, col sep=comma] {./Data/RICE_USTM_INDUCED1x1_6dB_kappa_1000_n_168.csv}coordinate[pos=0.07](ut1);      
        \addplot[name path = p2,color = red, solid,mark=*,mark size=\ptsize,forget plot] table [y index={2}, col sep=comma] {./Data/RICE_USTM_INDUCED1x1_6dB_kappa_1000_n_168.csv}coordinate[pos=0.07](pt1);     
		\addplot[color=gray, opacity=0.2] fill between[of=p1 and p2];
		
		\coordinate (pt2) at ($(pt1) !.5! (ut1)$);
		\draw (pt2) ellipse  (3pt and 5pt);
		\coordinate (pt3) at ($(pt2)+ (0pt,-5pt)$);
		\coordinate (pt4) at ($(pt3)+ (+5pt,-14pt)$);
		\draw[<-] (pt3)--(pt4) node at ($(pt4) + (13pt,-2pt)$) {$\kappa=10^3$};		
		 \addplot[name path = p1,color = blue, solid,mark=*,mark size=\ptsize,forget plot]  table [y index={1}, col sep=comma] {./Data/RICE_USTM_INDUCED1x1_6dB_kappa_100_n_168.csv}coordinate[pos=0.03](ut1); 
        \addplot[name path = p2,color = red, solid,mark=*,mark size=\ptsize,forget plot] table [y index={2}, col sep=comma] {./Data/RICE_USTM_INDUCED1x1_6dB_kappa_100_n_168.csv}coordinate[pos=0.03](pt1);      
		\addplot[color=red, opacity=0.2] fill between[of=p1 and p2];
		
		\coordinate (pt2) at ($(pt1) !.5! (ut1)$);
		\draw (pt2) ellipse  (3pt and 5pt);
		\coordinate (pt3) at ($(pt2)+ (1pt,-5pt)$);
		\coordinate (pt4) at ($(pt3)+ (+5pt,-14pt)$);
		\draw[<-] (pt3)--(pt4) node at ($(pt4) + (11pt,-2pt)$) {$\kappa=10^2$};		
			\addplot[name path = p1,color = blue, solid,mark=*,mark size=\ptsize,forget plot] table [y index={1}, col sep=comma] {./Data/RICE_USTM_INDUCED1x1_6dB_kappa_10_n_168.csv}coordinate[pos=0.05](ut1);     
        \addplot[name path = p2,color = red, solid,mark=*,mark size=\ptsize,forget plot] table [y index={2}, col sep=comma] {./Data/RICE_USTM_INDUCED1x1_6dB_kappa_10_n_168.csv}coordinate[pos=0.05](pt1);      
		\addplot[color=blue, opacity=0.2] fill between[of=p1 and p2];   
		
		\coordinate (pt2) at ($(pt1) !.5! (ut1)$);
		\draw (pt2) ellipse  (3pt and 5pt);
		\coordinate (pt3) at ($(pt2)+ (0pt,-5pt)$);
		\coordinate (pt4) at ($(pt3)+ (+5pt,-14pt)$);
		\draw[<-] (pt3)--(pt4) node at ($(pt4) + (11pt,-2pt)$) {$\kappa=10$};			
		\addplot[name path = p1, color = blue, solid,mark=*,mark size=\ptsize,forget plot] table [y index={1}, col sep=comma] {./Data/RICE_USTM_INDUCED1x1_6dB_kappa_1_n_168.csv}coordinate[pos=0.78](ut1);     
        \addplot[name path = p2, color = red, solid,mark=*,mark size=\ptsize,forget plot] table [y index={2}, col sep=comma] {./Data/RICE_USTM_INDUCED1x1_6dB_kappa_1_n_168.csv}coordinate[pos=0.78](pt1);         	
		\addplot[color=green, opacity=0.2] fill between[of=p1 and p2];
		
		\coordinate (pt2) at ($(pt1) !.5! (ut1)$);
		\draw (pt2) ellipse  (3pt and 5pt);
		\coordinate (pt3) at ($(pt2)+ (0pt,5pt)$);
		\coordinate (pt4) at ($(pt3)+ (+5pt,14pt)$);
		\draw[<-] (pt3)--(pt4) node[anchor=south ] {$\kappa=1$};			
		\addplot[name path = p1, color = blue, solid,mark=*,mark size=\ptsize,forget plot] table [y index={1}, col sep=comma] {./Data/RICE_USTM_INDUCED1x1_6dB_kappa_0_n_168.csv}coordinate[pos=0.92](ut1);   
        \addplot[name path = p2, color = red, solid,mark=*,mark size=\ptsize,forget plot] table [y index={2}, col sep=comma] {./Data/RICE_USTM_INDUCED1x1_6dB_kappa_0_n_168.csv}coordinate[pos=0.902](pt1);          	
		\addplot[color=yellow, opacity=0.2] fill between[of=p1 and p2];
		
		\coordinate (pt2) at ($(pt1) !.5! (ut1)$);
		\draw (pt2) ellipse  (3pt and 5pt);
		\coordinate (pt3) at ($(pt2)+ (0pt,-5pt)$);
		\coordinate (pt4) at ($(pt3)+ (+5pt,-14pt)$);
		\draw[<-] (pt3)--(pt4) node[anchor=north ] {$\kappa=0$};			
        \end{axis}
        \begin{axis}
[
      xmin = 2,
      xmax = 84,
 	  xlabel = Size of coherence block $n_c$,
      xmode = log, 
      xticklabels={{$84$}, {$42$}, {$24$}, {$12$}, {$8$}, {$6$},{$4$},{$2$}},
      xtick={2,  4, 7, 14,  21, 28, 42,84},   
      hide y axis,
      axis x line*=top,
      xlabel near ticks
]
    \addplot[draw=none]{0};
    \end{axis}
    \end{tikzpicture}
    \caption{Achievability (red) and converse (blue) bounds on \Rmax from Theorem~\ref{thm:DT} and~\ref{thm:MC}, respectively. Here, $\kappa=\lrb{0,1,10,100,1000}$, $n = 168$, $\err=10^{-3}$, and $\rho = 6$ dB.}
     \label{fig:1}
\end{figure}
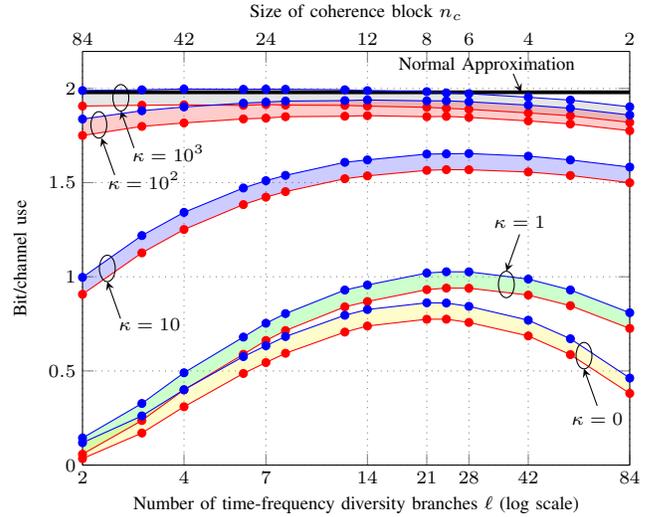

In Fig. \ref{fig:1}, we plot the  bounds on \Rmax given in  Theorem \ref{thm:DT} and \ref{thm:MC} for different values of the Rician factor $\kappa$.
We assume a blocklength of $n=168$ channel uses; furthermore, $\epsilon=10^{-3}$ and $\rho=6\dB$. The bounds are depicted as a function of the number of time-frequency diversity branches $\ell$ or, equivalently, the size of each coherence block $\nc$. 
We see from Fig.~\ref{fig:1} that there exists an optimal number of diversity branches that maximizes \Rmax. 
When $\ell$ is too low, the performance bottleneck is the limited diversity available. When $\ell$ is too high, the limiting factor is instead the fast variation of the channel.
We note also that \Rmax increases with $\kappa$ and it  becomes less sensitive to $\ell$ as $\kappa$ grows.
This is expected since when $\kappa\to \infty$ the Rician channel becomes an AWGN channel.
Indeed, we see that the bounds obtained for the case $\kappa=10^{3}$ are in good agreement with the normal approximation~\eqref{eq:normal_approximation}.

\subsection{Pilot-Assisted Transmission} 
\label{sec:numerical_results_b}
\begin{figure}[h]
 \centering
\begin{tikzpicture}
\begin{axis}
    [
    	axis x line*=bottom,
    	xmode = log,        
         xlabel={Number of time-frequency diversity branches $\ell$ (log scale)},
        ylabel = Bit/channel use,
        xmin = 2,
        xmax = 84,
        ymin = 0,
        xticklabels={{$2$}, {$4$}, {$7$}, {$14$}, {$21$},{$28$}, {$42$},{$84$}},
       xtick={2, 4 , 7,14,21,28,42, 84},   
		legend style={at={(axis cs:5,0)},anchor=south west},
		 xlabel near ticks,
		 grid=both
    ]
		 \def\ptsize{1.5}
      
    \addplot[name path = p1,color = blue, solid,mark=*,mark size=\ptsize,forget plot] table [y index={1}, col sep=comma] {./Data/RICE_USTM_INDUCED1x1_6dB_kappa_10_n_168.csv}coordinate[pos=0.95](ut1);  
    \addplot[name path = p2,color = red, solid,mark=*,mark size=\ptsize,forget plot] table [y index={2}, col sep=comma] {./Data/RICE_USTM_INDUCED1x1_6dB_kappa_10_n_168.csv}coordinate[pos=0.95](pt1);      
	\addplot[color=gray, opacity=0.2] fill between[of=p1 and p2];
	
		\coordinate (pt2) at ($(pt1) !.5! (ut1)$);
		\draw (pt2) ellipse  (3pt and 5pt);
		\coordinate (pt3) at ($(pt2)+ (0pt,-5pt)$);
		\coordinate (pt4) at ($(pt3)+ (0pt,-9pt)$);
		\draw[<-] (pt3)--(pt4) node at ($(pt4) + (-3pt,-3pt)$){$\np=0$};			
 
	    \addplot[name path = p1,color = orange, solid,mark=*,mark size=\ptsize,forget plot] table [y index={2}, col sep=comma] {./Data/RICE_PAT_1x1_6dB_kappa_10_n_168_p1.csv}coordinate[pos=0.20](p1);
		\coordinate (p2) at ($(p1)+ (7pt,0pt)$); \draw[<-] (p1)--(p2) node[anchor= west] {$\np = 1$};

		\addplot[name path = p1,color = magenta, solid,mark=*,mark size=\ptsize,forget plot] table [y index={2}, col sep=comma] {./Data/RICE_PAT_1x1_6dB_kappa_10_n_168_p2.csv}coordinate[pos=0.25](p1);     
		\coordinate (p2) at ($(p1)+ (6pt,0pt)$); \draw[<-] (p1)--(p2) node[anchor= west] {$\np = 2$};

		\addplot[name path = p1,color = green, solid,mark=*,mark size=\ptsize,forget plot] table [y index={2}, col sep=comma] {./Data/RICE_PAT_1x1_6dB_kappa_10_n_168_p4.csv}coordinate[pos=0.50](p1);    
		\coordinate (p2) at ($(p1)+ (6pt,0pt)$); \draw[<-] (p1)--(p2) node[anchor= west] {$\np = 4$};
	
		\addplot[name path = p1,color = purple, solid,mark=*,mark size=\ptsize,forget plot] table [y index={2}, col sep=comma] {./Data/RICE_PAT_1x1_6dB_kappa_10_n_168_p6.csv}coordinate[pos=0.44](p1);    
		\coordinate (p2) at ($(p1)+ (-20pt,0pt)$); \draw[<-] (p1)--(p2) node[anchor= east] {$\np = 6$};

		 \addplot[name path = p1,color = black, solid,mark=*,mark size=\ptsize,forget plot] table [y index={2}, col sep=comma] {./Data/RICE_PAT_1x1_6dB_kappa_10_n_168_p8.csv}coordinate[pos=0.41](p1);  
		\coordinate (p2) at ($(p1)+ (-8pt,0pt)$); \draw[<-] (p1)--(p2) node[anchor= east] {$\np = 8$};

        \end{axis}
         \begin{axis}
[
      xmin = 2,
      xmax = 84,
 	  xlabel = Size of coherence block $n_c$,
      xmode = log, 
      xticklabels={{$84$}, {$42$}, {$24$}, {$12$}, {$8$}, {$6$},{$4$},{$2$}},
      xtick={2,  4, 7, 14,  21, 28, 42,84},   
      hide y axis,
      axis x line*=top,
      xlabel near ticks
]
    \addplot[draw=none]{0};
    \end{axis}
    \end{tikzpicture}
        \caption{Comparison between the converse bound (blue) given in Theorem~\ref{thm:MC} and the achievability bound with pilot-assisted transmission given in Theorem~\ref{thm:pilot} for the case when $\np = \lrb{0,1,2,4,6,8}$ pilot symbols are inserted within each coherence block. Here, $\kappa=10$, $n = 168$, $\err=10^{-3}$ and $\rho = 6$~dB.}
     \label{fig:3}
\end{figure}
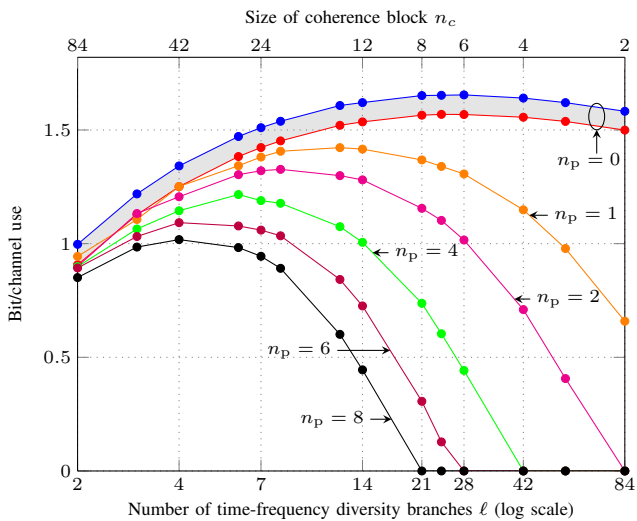

In Fig.~\ref{fig:2}, we compare the pilot-assisted-transmission achievability bound given in Theorem~\ref{thm:pilot} with  $\np\in\{0,1,2,4,6,8\}$, with the converse bound given in Theorem~\ref{thm:MC}. 
We assume $\kappa=0$.
The other parameters are set as in Fig.~\ref{fig:1}.
We can see from  Fig.~\ref{fig:2} that using one pilot yields similar performance as using the noncoherent shell-code scheme in Theorem~\ref{thm:DT}.
Transmitting more than one pilot turns out to be detrimental when the size of the coherence block decreases. 
Indeed, the improvement in the channel estimate is outweighed by the rate loss caused by pilot insertion.
As shown in Fig.~\ref{fig:3} for the case $\kappa=10$, the negative effect of pilot overhead becomes more significant when $\kappa$ is large.%
%


 \begin{figure}[h]
 \centering
\begin{tikzpicture}
\begin{axis}
    [
    	axis x line*=bottom,
    	xmode = log,        
         xlabel={Number of time-frequency diversity branches $\ell$ (log scale)},
        ylabel = Bit/channel use,
        xmin = 2,
        xmax = 84,
        ymin = 0,
        xticklabels={{$2$}, {$4$}, {$7$}, {$14$}, {$21$},{$28$}, {$42$},{$84$}},
       xtick={2, 4 , 7,14,21,28,42, 84},   
		legend style={at={(axis cs:5,0)},anchor=south west},
		 xlabel near ticks,
		 grid=both
    ]
		 \def\ptsize{1.5}
      
    \addplot[name path = p1,color = blue, solid,mark=*,mark size=\ptsize,forget plot] table [y index={1}, col sep=comma] {./Data/RICE_USTM_INDUCED1x1_6dB_kappa_0_n_168.csv}coordinate[pos=0.95](ut1);  
    \addplot[name path = p2,color = red, solid,mark=*,mark size=\ptsize,forget plot] table [y index={2}, col sep=comma] {./Data/RICE_USTM_INDUCED1x1_6dB_kappa_0_n_168.csv}coordinate[pos=0.95](pt1);      
	\addplot[color=gray, opacity=0.2] fill between[of=p1 and p2];
	
		\coordinate (pt2) at ($(pt1) !.5! (ut1)$);
		\draw (pt2) ellipse  (3pt and 7pt);
		\coordinate (pt3) at ($(pt2)+ (0pt,+7pt)$);
		\coordinate (pt4) at ($(pt3)+ (0pt,+30pt)$);
		\draw[<-] (pt3)--(pt4) node at ($(pt4) + (-3pt,3pt)$){$\np=0$};			
 
	    \addplot[name path = p1,color = orange, solid,mark=*,mark size=\ptsize,forget plot] table [y index={2}, col sep=comma] {./Data/RICE_PAT_1x1_6dB_kappa_0_n_168_p1.csv}coordinate[pos=0.02](p1);
		\coordinate (p2) at ($(p1)+ (-32pt,0pt)$); \draw[<-] (p1)--(p2) node at ($(p2) + (-14pt,0pt)$){$\np=1$};

		\addplot[name path = p1,color = magenta, solid,mark=*,mark size=\ptsize,forget plot] table [y index={2}, col sep=comma] {./Data/RICE_PAT_1x1_6dB_kappa_0_n_168_p2.csv}coordinate[pos=0.10](p1);     
		\coordinate (p2) at ($(p1)+ (-6pt,0pt)$); \draw[<-] (p1)--(p2) node[anchor= east] {$\np = 2$};

		\addplot[name path = p1,color = green, solid,mark=*,mark size=\ptsize,forget plot] table [y index={2}, col sep=comma] {./Data/RICE_PAT_1x1_6dB_kappa_0_n_168_p4.csv}coordinate[pos=0.45](p1);    
		\coordinate (p2) at ($(p1)+ (6pt,0pt)$); \draw[<-] (p1)--(p2) node[anchor= west] {$\np = 4$};
	
		\addplot[name path = p1,color = purple, solid,mark=*,mark size=\ptsize,forget plot] table [y index={2}, col sep=comma] {./Data/RICE_PAT_1x1_6dB_kappa_0_n_168_p6.csv}coordinate[pos=0.38](p1);    
		\coordinate (p2) at ($(p1)+ (-20pt,0pt)$); \draw[<-] (p1)--(p2) node[anchor= east] {$\np = 6$};

		 \addplot[name path = p1,color = black, solid,mark=*,mark size=\ptsize,forget plot] table [y index={2}, col sep=comma] {./Data/RICE_PAT_1x1_6dB_kappa_0_n_168_p8.csv}coordinate[pos=0.41](p1);  
		\coordinate (p2) at ($(p1)+ (-8pt,0pt)$); \draw[<-] (p1)--(p2) node[anchor= east] {$\np = 8$};

        \end{axis}
         \begin{axis}
[
      xmin = 2,
      xmax = 84,
 	  xlabel = Size of coherence block $n_c$,
      xmode = log, 
      xticklabels={{$84$}, {$42$}, {$24$}, {$12$}, {$8$}, {$6$},{$4$},{$2$}},
      xtick={2,  4, 7, 14,  21, 28, 42,84},   
      hide y axis,
      axis x line*=top,
      xlabel near ticks
]
    \addplot[draw=none]{0};
    \end{axis}
    \end{tikzpicture}
   
        \caption{Comparison between the converse bound (blue) given in Theorem~\ref{thm:MC} and the achievability bound with pilot-assisted transmission given in Theorem~\ref{thm:pilot} for the case when $\np = \lrb{0,1,2,4,6,8}$ pilot symbols are inserted within each coherence block. Here, $\kappa=0$, $n = 168$, $\err=10^{-3}$ and $\rho = 6$~dB.}
 \label{fig:2}
\end{figure}
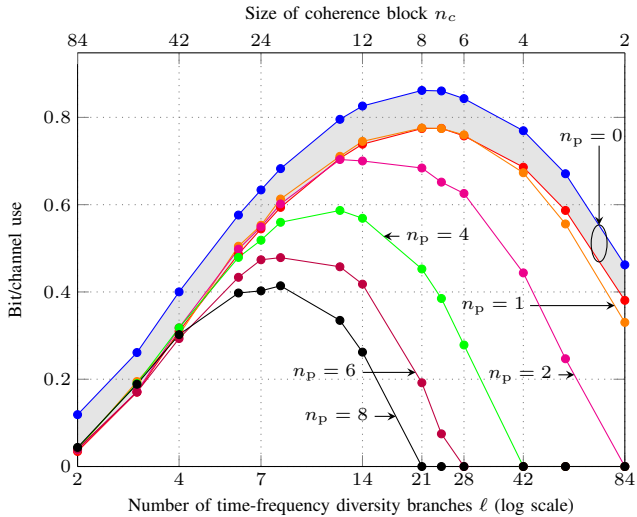

\subsection{Conclusion}
We presented finite-blocklength bounds on the maximum coding rate achievable over Rician block-fading channels for the case when no \emph{a priori} CSI is available.
Our bounds allow one to estimate the optimal number of time-frequency branches over which one should code across. 
This value trades optimally  the rate gains resulting from time-frequency diversity against the rate loss resulting from fast channel variations.
We also obtained an achievability bound for the case of pilot-assisted transmission, which allow one to optimize the number of pilot symbols to be transmitted within each coherence block.
Our results indicate that pilot-assisted transmission results in a significant rate loss when the coherence block is short and when the Rician factor $\kappa$ is large. 
In these situations, noncoherent transmission schemes are preferable.
A comparison between our bounds and the performance of actual coding schemes, along the lines of what we recently reported in~\cite{ostman17-02a}, is left for future work.

\bibliographystyle{IEEEtran}
\bibliography{./giubib}
\end{document}

%% file: symbols.tex
\DeclareSymbolFontAlphabet{\amsmathbb}{AMSb}%

\newcommand{\lro}[1]{\lefto({#1}\right)}																
\newcommand{\lrbo}[1]{\lefto \lbrace {#1} \right \rbrace}															
\newcommand{\lrho}[1]{\lefto [ {#1} \right ]}																				

\newcommand{\lr}[1]{\left({#1}\right)}																
\newcommand{\lrb}[1]{\left \lbrace {#1} \right \rbrace}															
\newcommand{\lrh}[1]{\left [ {#1} \right ]}																				

\safemath{\dopplerspread}{B_D}																								
\safemath{\delayspread}{T_D}																									
\safemath{\nc}{n\sub{c}}																										
\safemath{\nd}{n\sub{d}}																										
\safemath{\ntx}{n\sub{t}} 																											
\safemath{\nrx}{n\sub{r}}																											
\safemath{\ntxt}{\tilde{n\sub{t}}}																											
\safemath{\cb}{\ensuremath{L}} 																								
\safemath{\cl}{\ensuremath{n}} 																								
\safemath{\txanto}{{\ensuremath{\tilde{m}_t}}} 																		
\safemath{\cs}{M} 																														
\safemath{\idPustm}{\ensuremath{S_{k}}}
\safemath{\error}{\ensuremath{\epsilon}} 																				
\safemath{\eexp}{\ensuremath{\mathcal{E}}} 																			
\safemath{\nsubc}{n\sub{s}}			 																						
\safemath{\nofdm}{n\sub{o}} 																									
\safemath{\bc}{\ensuremath{B_c}} 																							
\safemath{\ts}{\ensuremath{T_s}} 																							
\safemath{\nrb}{\ensuremath{n_{rb}}} 																						
\safemath{\nres}{\ell}
\newcommand{\cgauss}[2]{\mathcal{CN}\lro{\ensuremath{#1, #2}  }}   								
\safemath{\maxk}{M^*\lr{\nres, \nsubc, \nofdm, \epsilon, \rho}}
\safemath{\Rmax}{R^*}
\safemath{\np}{\ensuremath{n\sub{p}}}
\safemath{\code}{\ensuremath{\mathcal{C}}}
\safemath{\err}{\ensuremath{\epsilon}}

\safemath{\mI}{\ensuremath{i\lro{\randvecy ; \randvecx}}} 				

\safemath{\fyx}{\ensuremath{f_{\randvecy \given \randvecx = \vecx}\lro{\vecy}}} 			
\safemath{\fy}{\ensuremath{f_{\randvecy }\lro{\vecy}}} 										
\safemath{\fyb}{\ensuremath{f_{\randvecy }\lro{\bar{\vecy}}}} 										
\safemath{\fyhh}{\ensuremath{f_{\randvecy \given \hat{H}=\hat{h} }\lro{\vecy}}} 							
\safemath{\fyh}{\ensuremath{f_{\randvecy \given H=h}\lro{\vecy}}} 							
\safemath{\fh}{\ensuremath{f_{H }}\lro{h}} 																		
\safemath{\fhh}{\ensuremath{f_{\hat{H}\given H =h}\lro{\hat{h}}}} 															
\safemath{\fx}{\ensuremath{f_{\randvecx }\lro{\vecx}}} 															
\safemath{\fyxh}{\ensuremath{f_{\randvecy \given \randvecx=\vecx, H = h}\lro{\vecy}}} 															
\safemath{\fyxhh}{\ensuremath{f_{\randvecy \given \randvecx = \vecx, \hat{H} = \hat{h}}\lro{\vecy}}} 															
\safemath{\fyxl}{\ensuremath{f_{\randvecy_\ell \given \randvecx_\ell = \vecx_\ell }\lro{\vecy}}} 															
\safemath{\fyl}{\ensuremath{f_{\randvecy_\ell }\lro{\vecy}}} 															
\safemath{\fabsh}{\ensuremath{f_{\abs{ H }^2}\!\lr{z}}}

\safemath{\fyxr}{\ensuremath{f_{\randvecy \given \randvecx = \vecx}\lro{\randvecy}}} 			
\safemath{\fyhr}{\ensuremath{f_{\randvecy \given H=h}\lro{\randvecy}}} 			
\safemath{\fyxhr}{\ensuremath{f_{\randvecy \given \randvecx=\vecx, H=h}\lro{\randvecy}}} 			
\safemath{\fyr}{\ensuremath{f_{\randvecy }\lro{\randvecy}}} 										
\safemath{\fyxlr}{\ensuremath{f_{\randvecy_\ell\given \randvecx_\ell = \vecx_\ell }\lro{\randvecy}}} 															
\safemath{\fylr}{\ensuremath{f_{\randvecy_\ell}\lro{\randvecy}}} 															

\safemath{\pyx}{\ensuremath{P_{\randvecy \given \randvecx}}} 			
\safemath{\qy}{\ensuremath{Q_{\randvecy }}} 										
\safemath{\py}{\ensuremath{P_{\randvecy }}} 										
\safemath{\pyhh}{\ensuremath{P_{\randvecy \given \hat{H} }}} 							
\safemath{\pyxhh}{\ensuremath{P_{\randvecy \given \randvecx, \hat{H} }}} 							
\safemath{\pyxh}{\ensuremath{P_{\randvecy \given \randvecx, H}}} 							
\safemath{\ph}{\ensuremath{P_{H }}} 																		
\safemath{\phh}{\ensuremath{P_{\hat{H}\given H }}} 															
\safemath{\px}{\ensuremath{P_{\randvecx }}} 															

\safemath{\randveca}{\bm{A}}
\safemath{\randvecb}{\bm{B}}
\safemath{\randvecc}{\bm{C}}
\safemath{\randvecd}{\bm{D}}
\safemath{\randvece}{\bm{E}}
\safemath{\randvecf}{\bm{F}}
\safemath{\randvecg}{\bm{G}}
\safemath{\randvech}{\bm{H}}
\safemath{\randveci}{\bm{I}}
\safemath{\randvecj}{\bm{J}}
\safemath{\randveck}{\bm{K}}
\safemath{\randvecl}{\bm{L}}
\safemath{\randvecm}{\bm{M}}
\safemath{\randvecn}{\bm{N}}
\safemath{\randveco}{\bm{O}}
\safemath{\randvecp}{\bm{P}}
\safemath{\randvecq}{\bm{Q}}
\safemath{\randvecr}{\bm{R}}
\safemath{\randvecs}{\bm{S}}
\safemath{\randvect}{\bm{T}}
\safemath{\randvecu}{\bm{U}}
\safemath{\randvecv}{\bm{V}}
\safemath{\randvecw}{\bm{W}}
\safemath{\randvecx}{\bm{X}}
\safemath{\randvecy}{\bm{Y}}
\safemath{\randvecz}{\bm{Z}}
\safemath{\randvecphi}{\bm{\Phi}}

\safemath{\randmatA}{\amsmathbb{A}}
\safemath{\randmatB}{\amsmathbb{B}}
\safemath{\randmatC}{\amsmathbb{C}}
\safemath{\randmatD}{\amsmathbb{D}}
\safemath{\randmatE}{\amsmathbb{E}}
\safemath{\randmatF}{\amsmathbb{F}}
\safemath{\randmatG}{\amsmathbb{G}}
\safemath{\randmatH}{\amsmathbb{H}}
\safemath{\randmatI}{\amsmathbb{I}}
\safemath{\randmatJ}{\amsmathbb{J}}
\safemath{\randmatK}{\amsmathbb{K}}
\safemath{\randmatL}{\amsmathbb{L}}
\safemath{\randmatM}{\amsmathbb{M}}
\safemath{\randmatN}{\amsmathbb{N}}
\safemath{\randmatO}{\amsmathbb{O}}
\safemath{\randmatP}{\amsmathbb{P}}
\safemath{\randmatQ}{\amsmathbb{Q}}
\safemath{\randmatR}{\amsmathbb{R}}
\safemath{\randmatS}{\amsmathbb{S}}
\safemath{\randmatT}{\amsmathbb{T}}
\safemath{\randmatU}{\amsmathbb{U}}
\safemath{\randmatV}{\amsmathbb{V}}
\safemath{\randmatW}{\amsmathbb{W}}
\safemath{\randmatX}{\amsmathbb{X}}
\safemath{\randmatY}{\amsmathbb{Y}}
\safemath{\randmatZ}{\amsmathbb{Z}}
\safemath{\randmatSigma}{\mathbb{\Sigma}}
\safemath{\randmatPhi}{\mathbb{\Phi}}
\safemath{\randmatLambda}{\mathbb{\Lambda}}

\safemath{\matSigma}{\bm{\Sigma}}
\safemath{\matPhi}{\bm{\Phi}}
\safemath{\matLambda}{\bm{\Lambda}}